\documentstyle[a4wide,epsf,11pt,titlepage]{article}

\newcounter{nref}
\newcommand{\bbib}{%
  \renewcommand{\refname}{\large\bf References}%
  \begin{thebibliography}{99}%
  \setcounter{enumiv}{\thenref}}
\newcommand{\ebib}{%
  \end{thebibliography}%
  \setcounter{nref}{\arabic{enumiv}}}
\newcommand{\head}[3]{%
  \setcounter{nref}{0}%
  \thispagestyle{empty}%
  \section*{\LARGE\bf #1}%
  \stepcounter{section}%
  \addcontentsline{toc}{section}{#1}%
  \large\itshape%
  #2\\\vspace{0.1pt}\\%
  #3%
  \normalsize\upshape%
  \bigskip}
\begin{document}
\def\lsim{\mathrel{\rlap{\lower 4pt \hbox{\hskip 1pt $\sim$}}\raise 1pt \hbox
        {$<$}}}
\def\gsim{\mathrel{\rlap{\lower 4pt \hbox{\hskip 1pt $\sim$}}\raise 1pt \hbox
        {$>$}}}


\head{Pop III Hypernova Nucleosynthesis and Abundances in Very Metal-Poor
 Halo Stars}
     {Hideyuki\ Umeda$^1$, Ken'ichi Nomoto $^1$}
     {$^1$ Department of Astronomy, 
School of Science, The University of Tokyo, \\
7-3-1 Hongo, Bunkyo-ku, Tokyo 113-0033, Japan } 

\subsection*{Abstract}
   We calculate evolution and nucleosynthesis in 
massive Pop III stars with $M = 13 \sim 270M_\odot$, 
and compare the results with abundances of
very metal-poor halo stars.
The observed abundances can be explained by
the energetic core-collapse supernovae with $M \lsim 130M_\odot$
(``hypernovae'') but not by pair-instability
supernovae (PISNe) with $M \sim 140-270M_\odot$. 
This result constrain the IMF for the Pop III and
very metal-poor Pop II stars.

\subsection*{Observed Abundance and Hypernova Nucleosynthesis} 

The observed abundances of
metal-poor halo stars show quite interesting trends.  There are
significant differences between the abundance patterns in the iron-peak
elements below and above [Fe/H]$ \sim -2.5$.  
For [Fe/H]$\lsim -2.5$, the mean values of [Cr/Fe] and [Mn/Fe] 
decrease toward smaller metallicity, while [Co/Fe] and [Zn/Fe] 
increases (McWilliam et al. 1995; Ryan et al. 1996; Primas et al. 2000;
Blake et al. 2001; see also Fig.4).
The Galaxy was not well mixed in such a early stage, and the abundance
pattern of each supernova (SN) may be kept in the very metal-poor stars.
However, these trend could not be explained with the previous results of
nucleosynthesis by core-collapse SNe
such as Woosley \& Weaver (1995), Nomoto et al. (1997) and
Limongi et al. (2000).
Therefore, we need reconsideration for the nucleosynthesis of
very metal-poor stars.

\begin{figure}[h]
  \centerline{\epsfxsize=0.8\textwidth\epsffile{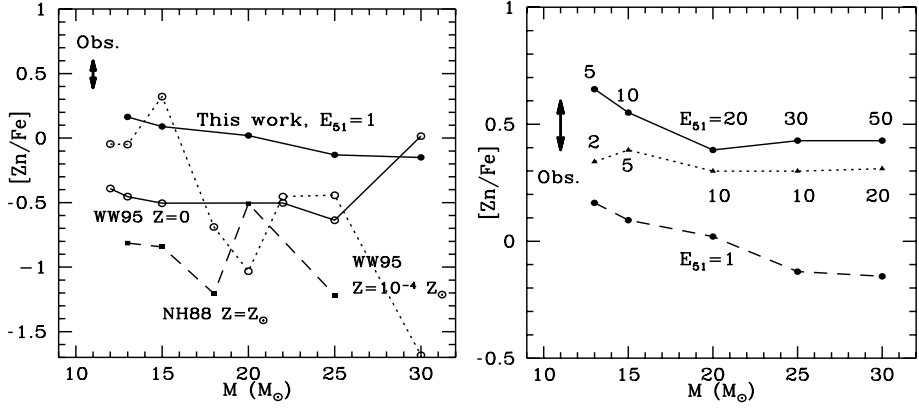}}
  \caption{Observed range of [Zn/Fe] for the very metal-poor 
stars ([Fe/H] $\lsim -3$) compared with the yields of theoretical
models with low explosion energy $E_{51}= E_{\rm exp}/10^{51}$ erg =1 (left)
and with high energies  $E_{51}= 2\sim50$ (right panel).
The observed large [Zn/Fe] value and also other unexpected
trends in [Co, Mn, Cr/Fe] can be explained with
this high energy (``Hypernova'') models.
  \label{umeda.fig1}}
\end{figure}

 We found that these trends are explained simultaneously in the context 
of Fe core-collapse SNe, if the mass from
complete Si-burning is relatively large compared with that from
incomplete Si-burning (Nakamura et al. 1999). However, all previous
model calculations including Nakamura et al. (1999)
underproduced the Zn/Fe ratio
significantly (Fig. 1 left).
In Umeda \& Nomoto (2002, UN02 hereafter), 
we find that [Zn/Fe] is larger for deeper
mass-cuts, smaller neutron excess, and larger explosion energies.  
Among them the large explosion energy is most important to realize
the large [Zn/Fe] ratio (Fig. 1 right).
The observed trends of the abundance ratios among the iron-peak elements
are better explained with
this high energy (``Hypernova'') models than the simple ``deep'' 
mass-cut effect, because 
the overabundance of Ni can be avoided in the hypernova models.

\subsection*{Pair Instability Supernovae}

\begin{figure}[h]
 \vskip -3.3cm
  \centerline{\epsfxsize=0.99\textwidth\epsffile{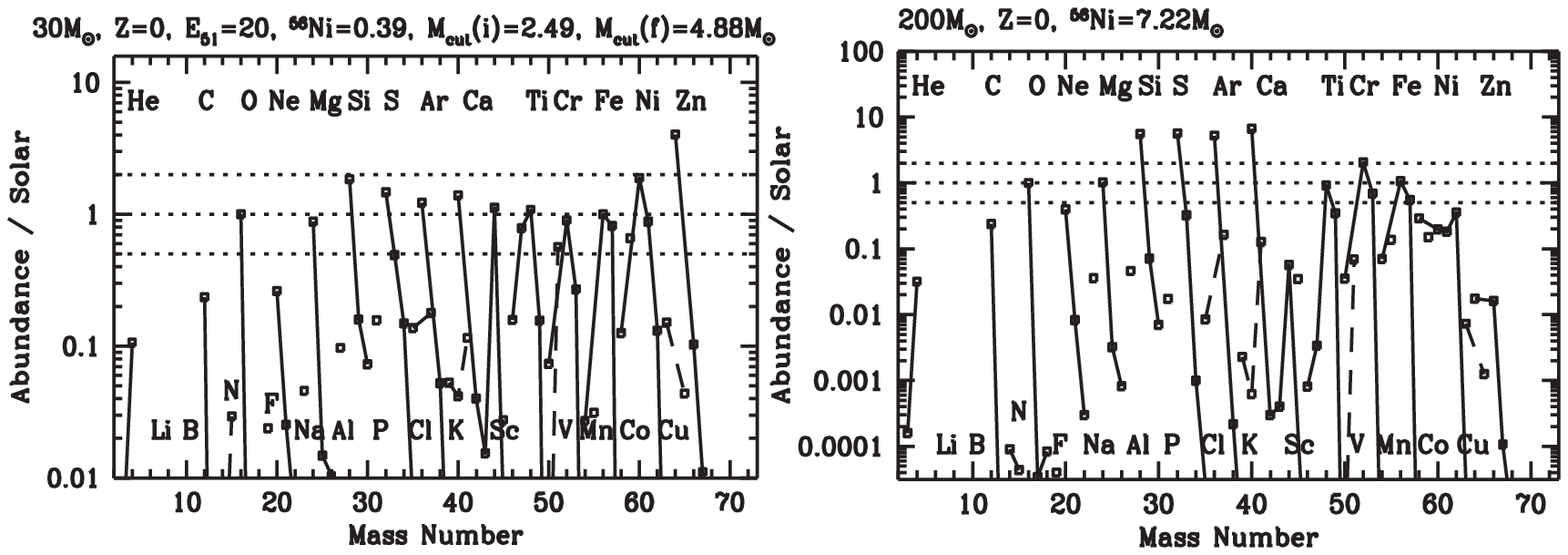}}
  \caption{The abundance pattern of a hypernova (left) and a 
PISN model (right panel), normalize to the solar abundance. 
For PISNe with initial masses $M \simeq 140 - 300 M_\odot$
 [Zn/Fe] is small, so that the abundance features of 
very metal-poor halo stars cannot be explained by these SNe.
  \label{umeda.fig2}}
\end{figure}

 We also investigate the yields of pair-instability supernova
explosions of $M \simeq 140 - 300 M_\odot$ stars. In Fig.2 
we compare the abundance pattern of a hypernova and a PISN model.
As can be seen [Zn/Fe] of PISNe are always small, because 
relatively large incomplete Si-burning region is formed
in the explosion
(UN02). Therefore,  
the abundance features of very metal-poor stars
cannot be explained by pair-instability supernovae.

\subsection*{[Fe/H] and SNe Mass, Energy}

 We have discussed that the large [Zn/Fe] ratio can be explained
with hypernova nucleosynthesis of $M \lsim 130M_\odot$ stars.
However, in order to explain the observed trend for [Fe/H], 
it is necessary to explain why hypernova contribution is 
dominant for lower [Fe/H]. Suppose that the star-formation
is induced by the supernova shock. Then the [Fe/H] of the
next generation stars is determined by the ratio of 
Fe mass synthesized by the SN
and the hydrogen mass swept by the supernova shock.
Since the latter is inversely proportional to the explosion
energy $E$, the following relation is obtained (Ryan et al. 1996;
Shigeyama \& Tsujimoto 1998):
\begin{equation}
 {\rm Fe/H} \propto {\rm M(^{56}Ni)}/E. 
\end{equation}

\begin{figure}[h]
\vskip -3.3cm
  \centerline{\epsfxsize=0.55\textwidth\epsffile{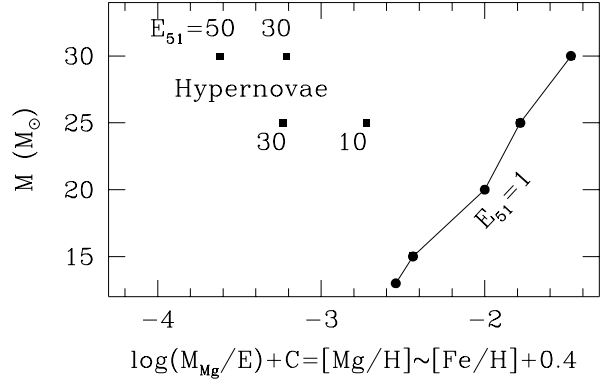}}
  \caption{ [(Mg, Fe)/H] determined by the relation (1) as a function of
SN mass and energy. The constant $C$ is
chosen to normalize the (20$M_\odot$, $E_{51}=1$) case to
be [Mg/H]= -2. 
  \label{umeda.fig3}}
\end{figure}

In Fig.3 we show the similar relation for Mg as a function of
the mass and energy of SNe. Although the ejected Fe mass is
theoretically uncertain, in most of the observed stars, Mg/Fe ratio
is roughly the same and typically [Mg/Fe] $\simeq$ 0.4.
Thus [Mg/H]$\sim$ [Fe/H]+0.4.
As shown in this figure, if the explosion energy $E_{51}$
is fixed, [(Mg, Fe)/H] increases with mass, because more Mg is
ejected for more massive stars. However the energy factor
is much more important, and hence
massive hypernova models
have smaller [(Mg, Fe)/H] than less massive normal SNe II. 
In this way, hypernova like nucleosynthesis patterns
can be seen in the very metal-poor stars.

\begin{figure}[h]
\vskip -0cm
   \centerline{\epsfxsize=0.4\textwidth\epsffile{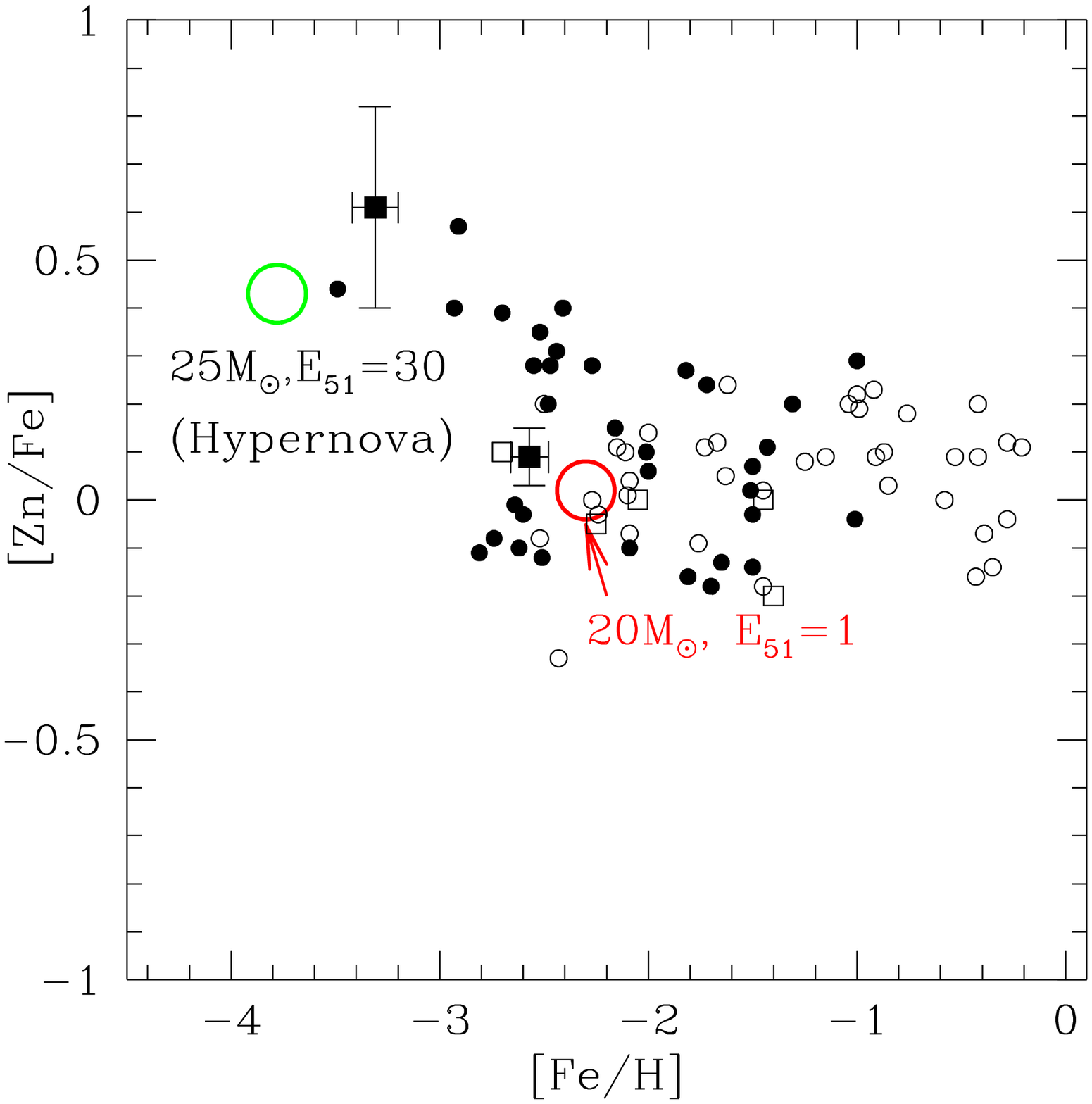}
               \epsfxsize=0.4\textwidth\epsffile{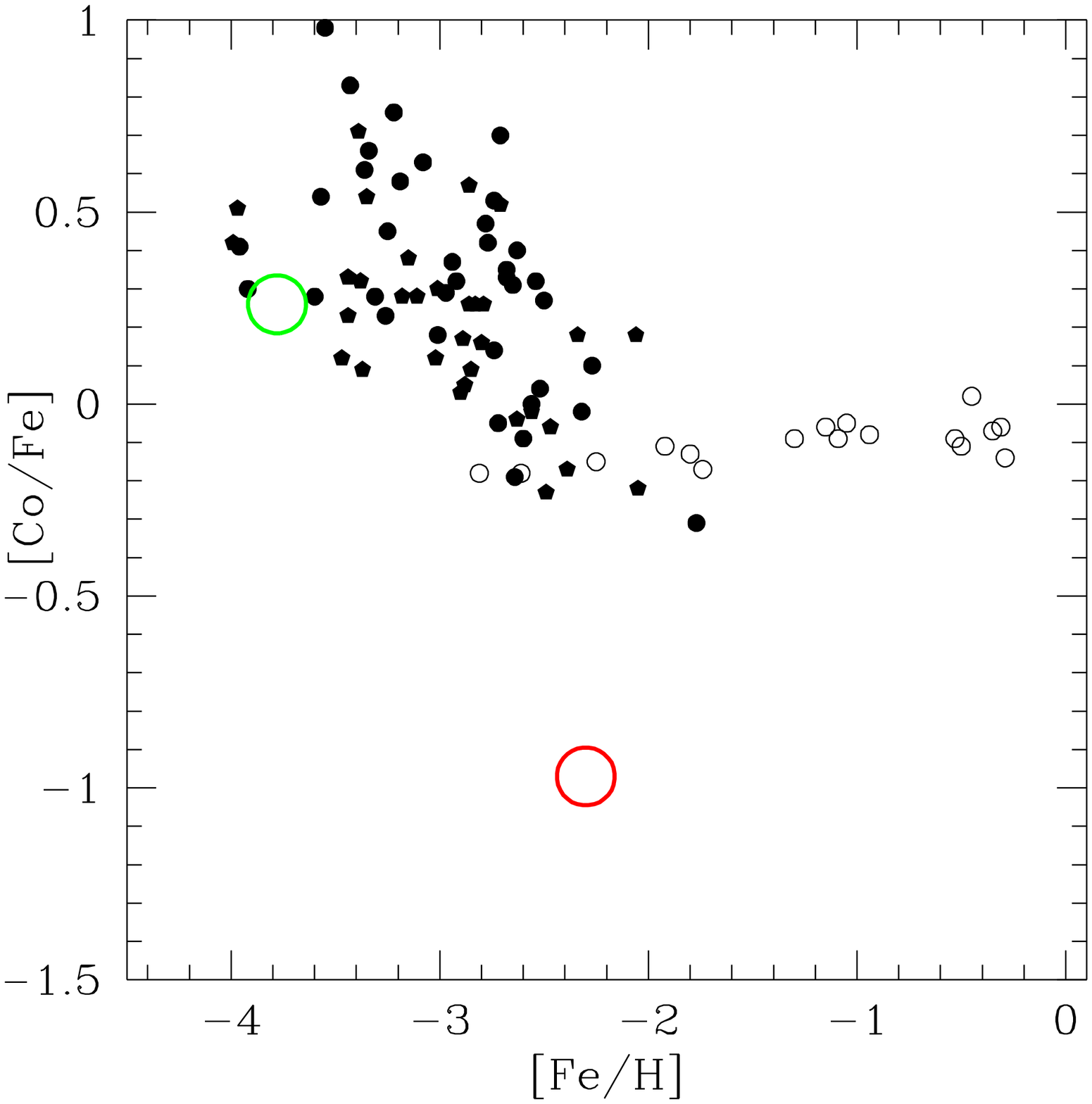}}
   \centerline{\epsfxsize=0.4\textwidth\epsffile{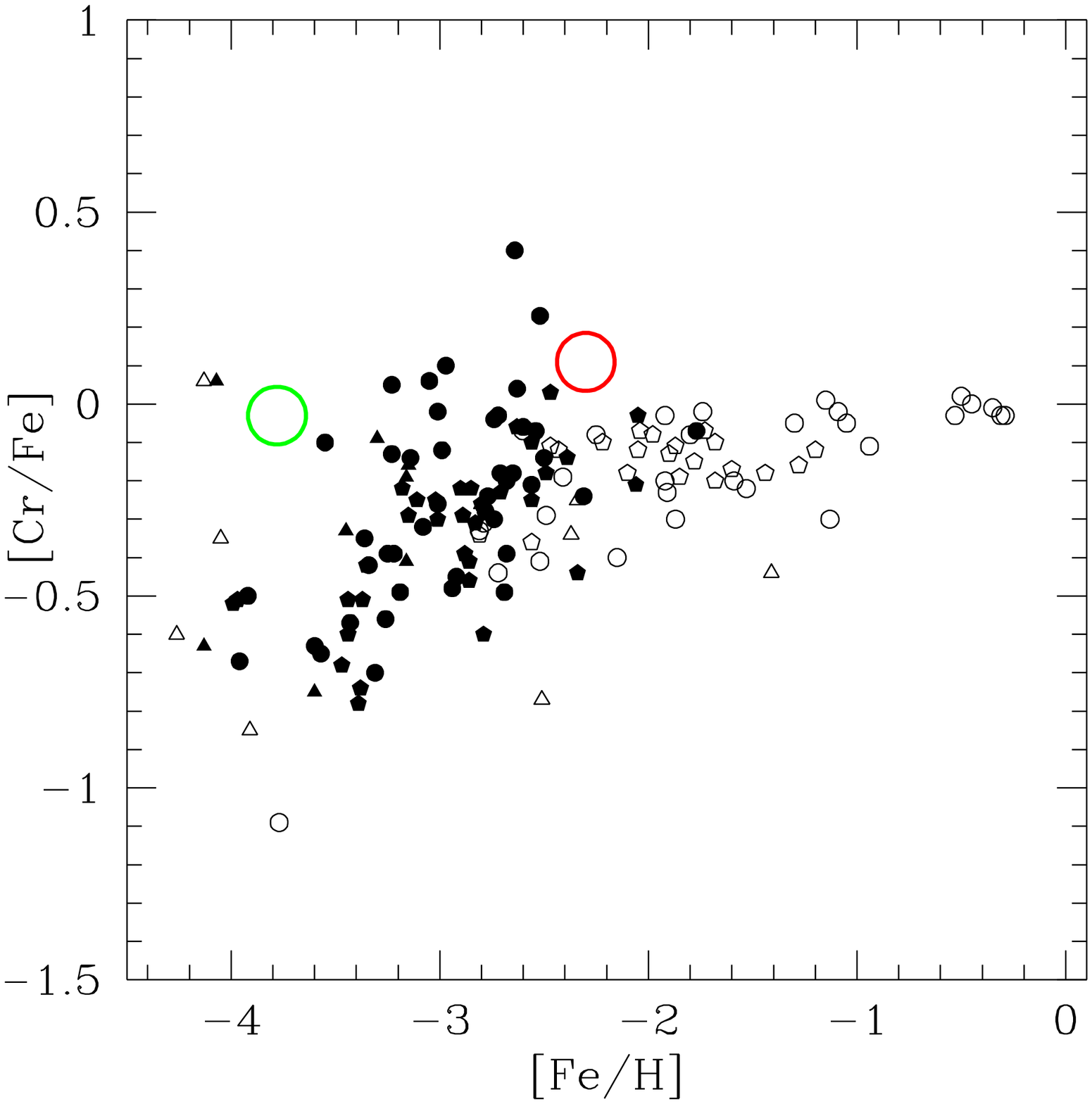}
               \epsfxsize=0.4\textwidth\epsffile{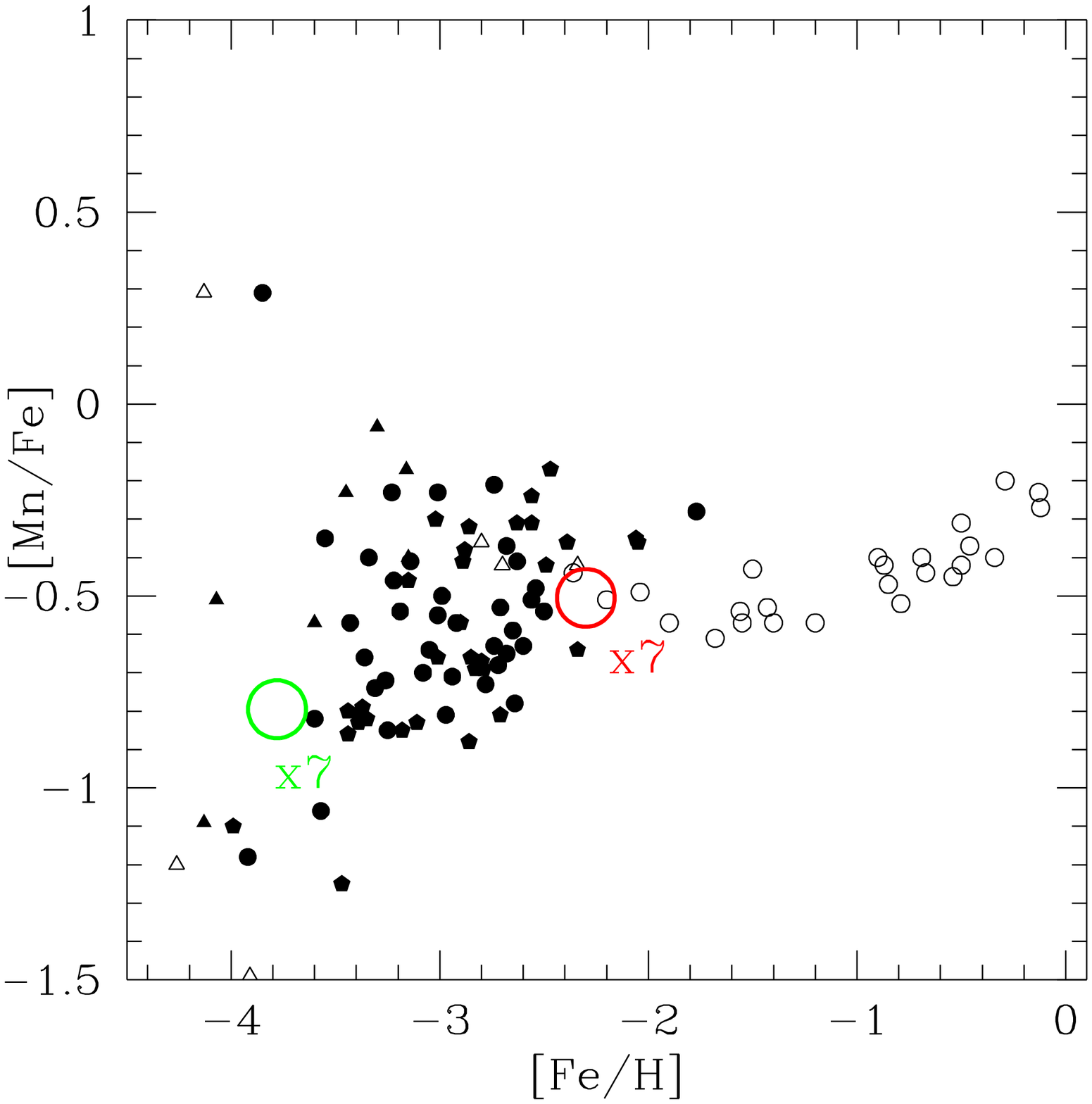}}
  \caption{Observational points of Fe-peak elements (small symbols)
and the theoretical abundance patterns for a
normal SNe II (20$M_\odot$, $E_{51}=1$) and a hypernova
(25$M_\odot$, $E_{51}=30$) models in UN2002 (large open circles).
Here, Mn abundance is multiplied by 7.  \label{umeda.fig4}}
\end{figure}

 Finally in Fig.4, together with the observation points,
we plot the abundance pattern of Fe-peak elements for a
normal SNe II (20$M_\odot$, $E_{51}=1$) and a hypernova
(25$M_\odot$, $E_{51}=30$) models in UN02.
Here the relative position in [Fe/H] is determined by the relation (1). 
Although there are some inconsistencies between the model
and observations, we can explain the trend in the observation.
As discussed in UN02, the Mn abundance is very sensitive
to $Y_{\rm e}$ of the progenitor, and thus the enhancement of factor
7 is not so difficult to realize. The remaining problem is
the overproduction of Cr and underproduction of Co,
which should be investigated further in future work.




\end{document}